\documentclass[prb,twocolumn,showpacs]{revtex4-1}
\usepackage{graphicx}
\usepackage{amsmath}
\usepackage{amssymb}

\newcommand{\ra}{\rangle}
\newcommand{\la}{\langle}
\DeclareMathAlphabet{\bi}{OML}{cmm}{b}{it}
\begin{document}

\title{In-plane electric field effect on a spin-orbit coupled two-dimensional 
electron system in presence of magnetic field}

\author{SK Firoz Islam and Tarun Kanti Ghosh}
\affiliation{Department of Physics, Indian Institute of Technology-Kanpur,
Kanpur-208 016, India}

\begin{abstract}
The effect of in-plane electric field on Landau level spacing, spin
splitting energy, average spin polarization and average spin current 
in the bulk as well as at the edges of a two-dimensional electron system 
with Rashba spin-orbit coupling are presented here. The spin splitting 
energy for a particular magnetic field is found to be reduced by the 
external in-plane electric field. Unlike the case of a
two-dimensional electron system without Rashba spin-orbit interaction, 
here the Landau level spacing is electric field dependent. This dependency 
becomes stronger at the edges in comparison to the bulk. 
The average spin polarization vector rotates anti-clockwise with the
increase of electric field. The average spin current also gets 
influenced significantly by the application of the in-plane electric 
field.

\end{abstract}



\date{\today}

\maketitle
\section{Introduction}
The proposal of the possible application of the spin degree of 
freedom in electronic devices has emerged as a new field called 
spintronics \cite{das1,appl1,appl2,appl3}. 
The experimental realization of the spin-orbit 
interaction which lifts the spin degeneracy even in absence of
magnetic field  has boosted research interest in this field.
The Rashba spin-orbit interaction (RSOI) is the particular interest 
of this field as it's strength can be enhanced significantly by 
applying a suitable gate voltage \cite{tech,matsu}. The origin of 
the RSOI is due to the lack of structural inversion symmetry in the 
semiconductor heterostructures \cite{rashba,rashba1}.
The RSOI provides a proficient way to control the electron's spin
degree of freedom. 
The electric field perpendicular to a two-dimensional electron system
(2DES) can enhance the Rashba spin-orbit coupling \cite{tech}.
The RSOI has manifested itself through some remarkable new features 
in different transport properties 
\cite{dutta,prl99,prl06,wang,wang1,firoz}.
Moreover, spin Hall effect (SHE) is one of the most important
consequences of the RSOI in a two-dimensional electron system (2DES). 
In SHE, electrons with opposite spin polarizations get scattered in 
transverse direction of the applied electric field even in absence 
of magnetic field. 
There have been several theoretical \cite{prl99,prl06,raimondi} as well as 
experimental \cite{kato,sinova} works on SHE.
The successful detection of spin accumulation \cite{kato,sinova} at the
edges has made the study of spin current interesting.

The effect of an uniform in-plane electric field on the Landau
levels in a two-dimensional system has been always an interesting 
quantum mechanical problem. 
The 2DES formed at the semiconductor heterostructure junction,
which is also called conventional 2DES and graphene monolayer are two
important examples of two-dimensional systems.
The spacing between two consecutive Landau levels of a conventional 
2DES is independent of electric field. 
On the other hand, the Landau level spacing in a graphene monolayer
depends on the transverse electric field and it can be collapsed 
under suitable strength of the electric field \cite{lukose,lukose1,ngu}.
These two results are obtained by exact analytical calculation.
The Landau levels of the Rashba spin-orbit coupled 2DES show two 
different energy branches with unequal level spacing \cite{wang}. 
The analytical 
derivation of Landau levels under in-plane electric field seems 
to be impossible. Therefore, we solve this problem by exact numerical
calculation.

The quantum Hall effect has been remained an attractive topic for
physicists. The edge states and edge charge currents play very 
important role in exploring details about quantum Hall effect 
\cite{halperin,macdonald,henrik}. 
The spatially well separated left- and right-moving edge states
give rise to non-dissipative transport \cite{muller,chang}.
In presence of the spin-orbit interaction, the transport of the
spin-polarized edge channels may provide rich physics. 
Several theoretical analysis of the spin edge states in a Rashba system 
has been also reported \cite{badalyan,rose}.
The crucial role played by the edge states in magnetization \cite{zhang} 
and spin polarization \cite{bao} in 2DES with RSOI has been studied 
extensively.

In this work, we present the effect of in-plane electric 
field on Landau level spacing (LLS), spin splitting energy, average 
spin polarization and average spin current in the bulk as well 
as at the edges numerically. We find the Landau level spacing and
the spin splitting energy decreases with increase of the electric 
field. The average spin polarization rotates anti-clockwise  
as we increase electric field. The average spin current is modified
due to the presence of the in-plane electric field.

This paper is organized as follows. In Sec. II, we formulate 
the numerical method to solve the Hamiltonian of a single 
electron confined in a finite size 2DES with RSOI in presence 
of crossed electric and magnetic fields. 
We also discuss the spin-split Landau levels, Landau level spacing,
average spin polarization and average spin current of an infinite 
2DES with RSOI in presence of magnetic field. In Section III, we present 
numerical results and discussion of the effect of
the transverse electric field on the spin splitting energy, LLS,
average spin polarization and average spin current.
However, as the energy becomes different at the
edges in comparison to the bulk for a finite system, the study of
above mentioned properties will also be covered 
at the edge with electric field. We present summary
of our work in Section IV.  

\section{Formalism for Numerical calculation}
We consider a spin-orbit coupled 2DES in a $xy$ plane of area 
$L_x \times L_y$ which is subjected to an in-plane electric filed 
$\vec{E}=E \hat{i}$ and magnetic field $\vec{B}=B\hat{z}$.
The single-electron Hamiltonian with electronic charge $q=-e$ is given by  
\begin{eqnarray} \label{hamil}
H & = & \Big[\frac{({\bf p} + e {\bf A})^2}{2m^*} + 
V(x) + eE x \Big] \sigma_0 \nonumber \\ 
& + &
\frac{\alpha}{\hbar}[\sigma_x (p_y + eA_y) - \sigma_y p_x]
+ \frac{g}{2}\mu_{B} B \sigma_z,
\end{eqnarray}
where $m^*$ is the effective mass of an electron, 
$\sigma_0$ is the $2 \times 2 $ identity matrix, 
$\sigma_i$ ($i=x,y,z$) are the Pauli matrices,
$\alpha$ is the Rashba spin-orbit coupling constant and 
$V(x)=0$ for $ |x| \leq L_x/2$ and otherwise infinity
is the hard-wall confining potential along the $x$ direction.
Also, $g$ is the effective Lande $g$-factor and 
$\mu_B = e\hbar/(2m_e)$ is the Bohr magneton with $m_e$ is the
free electron mass.
Here, we have chosen the Landau gauge $\vec{A} = B x \hat j$.

Before presenting numerical results, we shall review the
bulk Landau levels of a 2DES with RSOI in absence of the electric
field. The energy spectrum in the bulk can be obtained analytically by
diagonalizing the Hamiltonian \cite{wang}.
The RSOI mixes the two spin components.
For $s=0$, there is only one level, same as the lowest Landau
level without SOI, with energy
$ E_0^+ = E_0 = (\hbar \omega_c - g \mu_{B} B)/2 $ with
$ \omega_c = eB/m^*$.
The corresponding wave function is
\begin{equation}
\Psi_{0,k_y}^+({\bf r}) = \frac{e^{i k_y y}}{\sqrt{L_y}} \phi_0(x + x_0)
\left(
\begin{array}{c}
0 \\
1
\end{array}
\right),
\end{equation}
where $ x_0 = k_y l_0^2$ with magnetic length scale 
$l_0 = \sqrt{\hbar/(eB)}$.
For a given quantum number $s \geq 1$ there are two spin-split 
branches of energy levels, denoted by $+$ and $-$ with energies
\begin{equation}
E_s^{\pm} = s\hbar\omega_c{\pm} \sqrt{E_0^2 
+ s E_{\alpha} \hbar \omega_c},
\end{equation}
where $ E_{\alpha} = 2 m^{\ast} \alpha^2/\hbar^2$.
The corresponding wave function for $+$ branch is
\begin{equation}
\Psi _{s,k_y}^+({\bf r}) = \frac{e^{i k_y y}}{\sqrt{L_y A_s }} 
\left(\begin{array}{r} D_s \phi_{s-1}(x + x_0)
\\
\phi_s (x + x_0)
\end{array}
\right) \text{,}
\end{equation}
and for $-$ branch is
\begin{equation}
\Psi_{s,k_y}^-({\bf r})=\frac{e^{i k_y y}}{\sqrt{L_y A_s }} 
\left(\begin{array}{r} \phi_{s-1}(x + x_0)
\\
- D_s \phi_s(x + x_0)
\end{array}
\right) \text{,}
\end{equation}\\
where $A_s = 1 + D_s^2 $ with
$ D_s = \sqrt{s E_{\alpha} \hbar \omega_c}/
[E_0 +\sqrt{E_0^2 + s E_{\alpha} \hbar \omega_c}]$ and
$\phi_s(x) = (1/\sqrt{\sqrt{\pi }2^s s!l_0})
e^{-x^2/2l_0^2} H_s(x/l_0) $ is the normalized harmonic
oscillator wave function. Here,
$s$ is the Landau level index.

When both the hard-wall confining potential and the electric field 
are included, the Hamiltonian given by Eq. (\ref{hamil}) has to be 
solved numerically.
To solve the Hamiltonian numerically we choose the following 
wave function $\Psi (x,y)=(1/\sqrt{L_y})e^{ik_y y} \phi(x)$ with the 
function $\phi(x)$ expanded in the basis of the infinite potential 
well as
\begin{equation}
\phi(x) = \sqrt\frac{2}{L_x} \sum_n \sin \Big\{\frac{n \pi}{L_x} 
\Big(x+\frac{L_x}{2}\Big)\Big\}
\begin{bmatrix}a_n\\b_n \end{bmatrix}.
\end{equation}
Now, using the time-independent Schr\"odinger equation, 
we get the following matrix equations for the spinors:
\begin{eqnarray}
&\Big[&\frac{\hbar^2 \pi^2 l^2}{2m^* L_x^2} + 
\frac{g}{2}\mu_{B}\sigma_z B\Big]
\begin{bmatrix}a_l\\b_l\end{bmatrix}\nonumber\\
&+&\sum_n\Big[\sigma_x F_{ln} + i \sigma_y G_{ln} + M_{ln} + K_{ln} \Big]
\begin{bmatrix}a_n\\b_n\end{bmatrix}=0.
\end{eqnarray}
Here, the matrix elements are given by
\begin{eqnarray}
F_{ln} & = & \frac{2\alpha L_x}{(\pi l_0)^2}\int_{-\pi/2}^{\pi/2}  
\sin \Big\{l\Big(\theta + \frac{\pi}{2}\Big)\Big\}
(\theta + \theta_0) \nonumber \\ 
& \times &  \sin \Big\{n\Big(\theta+\frac{\pi}{2}\Big)\Big\}d\theta,
\end{eqnarray}
\begin{equation}
G_{ln}=\frac{2\alpha n}{L_x}\int_{-\pi/2}^{\pi/2}  
\sin \Big\{l\Big(\theta+\frac{\pi}{2}\Big)\Big\}
\cos \Big\{n\Big(\theta+\frac{\pi}{2}\Big)\Big\}d\theta,
\end{equation}
\begin{eqnarray}
M_{ln}&=&\frac{m^*\omega_c^2L_x^2}{\pi^3}\int_{-\pi/2}^{\pi/2} 
\sin \Big\{l\Big(\theta+\frac{\pi}{2}\Big)\Big\}
(\theta+\theta_0)^2\nonumber\\&\times&\sin 
\Big\{n\Big(\theta+\frac{\pi}{2}\Big)\Big\}d\theta 
\end{eqnarray}
and
\begin{equation}
K_{ln}=\frac{2eE L_x}{\pi^2}\int_{-\pi/2}^{\pi/2}  
\sin \Big\{l\Big(\theta+\frac{\pi}{2}\Big)\Big\}
\theta\sin \Big\{n\Big(\theta+\frac{\pi}{2}\Big)\Big\} d\theta,
\end{equation}
where $ \theta=\pi x/L_x$, $\theta_0=\pi x_0/L_x $.

We solve these equations numerically in a truncated Hilbert space
by considering $400 \times 400$ matrix Hamiltonian and
confirmed that first thirty eigenvalues in the bulk with ${\bf E } =0 $ 
exactly match with the results obtained from the analytical expression.
Also, the probability density of the corresponding low-energy states are 
in excellent agreement with the analytical results.
We have numerically checked that in presence of the electric field
$E$, the cyclotron center $x_0$ in the bulk is displaced by 
$ x_E = e E/(m^* \omega_c^2)$ from right to left side. 
For low-lying Landau levels, the 
electric field induced displacement $x_E$ is almost independent of $s$.
When $ x_0 \approx 0 $ the states are far away from the boundary 
and called bulk states. 
When $x_0  \approx \pm L_x/2$ the states are close to the boundary 
and called edge states.

We define the spin splitting energy for a given Landau 
level $s$ as $\Delta_s = E_s^{+} - E_s^{-}$.
In the bulk with $E=0$, 
$\Delta_s = 2 \sqrt{E_0^2 + sE_{\alpha} \hbar \omega_c}$.
It implies that the spin splitting energy $\Delta_s$ is
increasing with increase of $s$, $\alpha$ and $B$. 
The LLS between two consecutive Landau levels 
is defined as $\Delta E_s^{\pm}=E_{s+1}^{\pm}-E_{s}^{\pm}$.
The total of the LLS for spin-up and spin-down branches is
$\Delta E_s=\Delta E_s^{+} +\Delta E_s^{-}=2\hbar\omega_c$.
This is independent of $\alpha $ and Landau level index $s$.

The components of the average spin polarization are defined as
\begin{equation}
P_{s}^{\lambda, S_i}(k_y) = \frac{\hbar}{2} 
\int d^2r \Big[\Psi_{s,k_y}^{\lambda}({\bf r}) \Big]^{\dagger} 
\sigma_i \Psi_{s,k_y}^{\lambda}({\bf r}), 
\end{equation}
where $i =x,y,z$ and $\lambda = \pm $ represents the 
spin-up and spin-down components. 
The $y$ component of the average spin polarization is always zero.
In absence of the electric field, the average spin polarization 
components can be evaluated analytically. 
In the bulk, the average spin polarization of the 
ground state ($ s=0$) is along the +$z$ axis. In the excited states
($s \geq 1 $), the average spin polarization in the bulk is
$ P_{s}^{\lambda,S_z} = \lambda (\hbar/2)(1 - |D_s|^2)/A_s $. 
For a given $s$, the spin polarization of $+$ branch is anti-parallel 
to that of the $-$ branch.

The spin current density for a given Landau level $s$ is 
given by 
$ J_{s,y}^{\lambda, S_i}({\bf r}) = [\Psi_{s,y}^{\lambda}]^{\dagger}  
\hat{J}_{y}^{S_i} \Psi_{s,y}^{\lambda}$, 
where $\hat{J}_{y}^{S_i}$ are the 
spin current operators with $i=x,y,z$.
We use the conventional definition of the spin current operator as
$\hat{J}_{y}^{S_x} = \hbar \hat{v}_y \sigma_x/2$ and 
$\hat{J}_{y}^{S_z} = \hbar 
(\hat{v}_y\sigma_z + \sigma_z \hat{v}_y)/4$.
The $y$ component of the velocity operator is given by
$ \hat{v}_y=(x+x_0)\omega_c\sigma_0+\frac{\alpha}{\hbar}\sigma_x$.
The average spin current is defined as
$ \la J_{s,y}^{S_i} \ra = \sum_{\lambda} J_{s,y}^{\lambda, S_i}  
= \sum_{\lambda} \int d^2r
J_{s,y}^{\lambda, S_i}({\bf r}) $.
In absence of the electric field,  the average 
$x$-component spin current carried by a 
given Landau level can be easily obtained and it is given by
$\la J_{s,y}^{S_x}\ra=\alpha$, independent of the Landau level 
index $s$.
On the other hand, $z$ component of
average spin current is $\la J_{s,y}^{S_z}\ra=0$. 
A point to be noted here that average spin 
current is linearly dependent on $\alpha$ and independent of
magnetic field strength.
On the other hand, the total spin current in absence of magnetic field 
is cubic in $\alpha$ \cite{sonin}.
Following Ref.\cite{bao} , one can easily show that even in presence
of electric field the spin current $J_{s,y}^{\lambda, S_z} $
is proportional to $ P_{s}^{\lambda,S_x}$:
$ J_{s,y}^{\lambda, S_z} = - \frac{g \hbar \mu_B B}{2m^*}
P_{s}^{\lambda,S_x}$.
This kind of relation does not hold between
$ J_{s,y}^{\lambda, S_x} $ and $ P_{s}^{\lambda,S_z}$. 

\begin{figure}[t]
\begin{center}\leavevmode
\includegraphics[width=98mm]{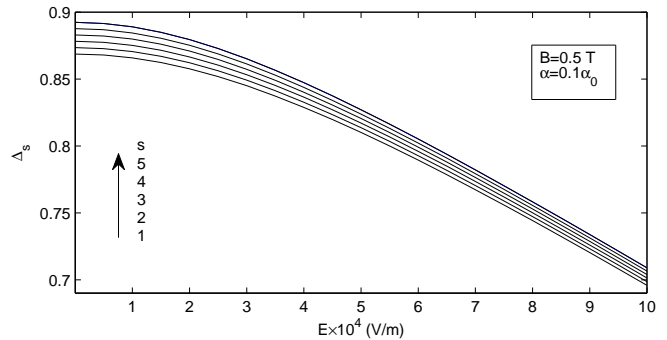}
\caption{Plots of the spin splitting energy in the bulk versus 
electric field for few $s$ values.}
\label{Fig1}
\end{center}
\end{figure}

\begin{figure}[t]
\begin{center}\leavevmode
\includegraphics[width=98mm]{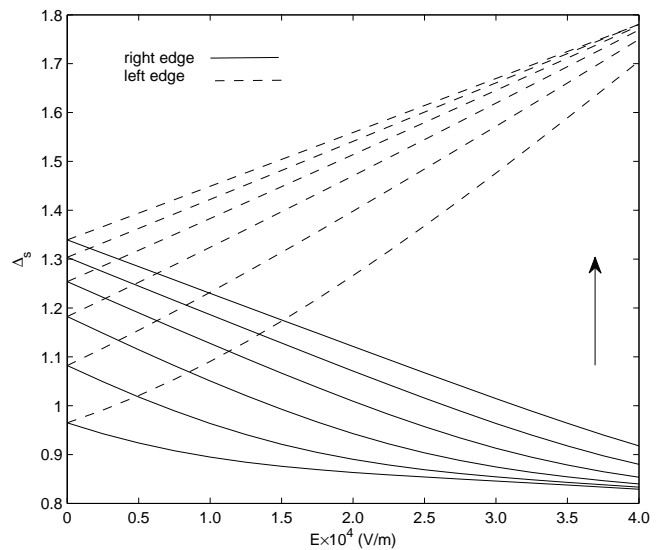}
\caption{Plots of the spin splitting energy at the edges versus 
electric field for few $s$ values.}
\label{Fig1}
\end{center}
\end{figure}

\begin{figure}[t]
\begin{center}\leavevmode
\includegraphics[width=98mm]{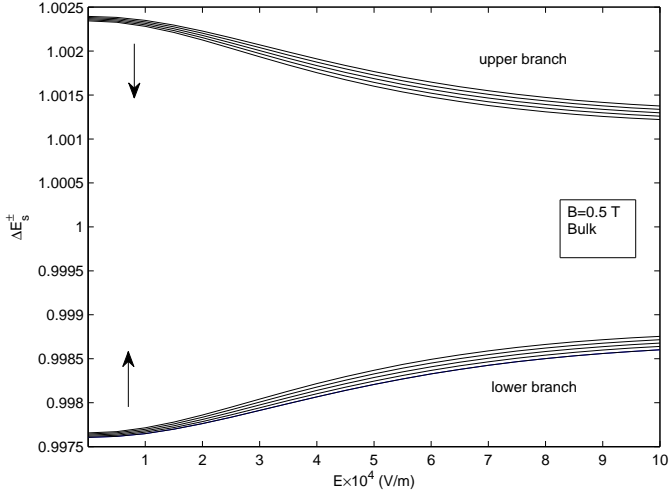}
\caption{Plots of the Landau level spacing in the bulk 
versus electric field for few $s$ values.}
\label{Fig2}
\end{center}
\end{figure}

\begin{figure}[t]
\begin{center}\leavevmode
\includegraphics[width=98mm]{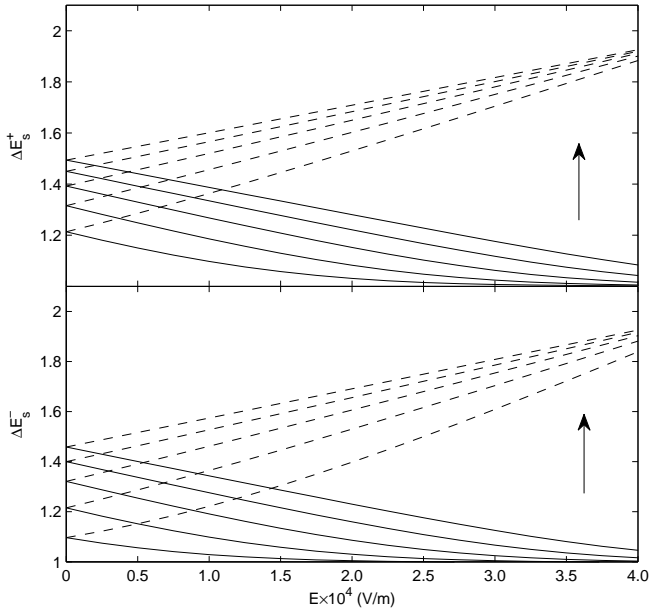}
\caption{Plots of the Landau level spacing  
versus electric field for few $s$ values. Here, dashed and
solid lines represent left and right edges.}
\label{Fig3}
\end{center}
\end{figure}

\begin{figure}[t]
\begin{center}\leavevmode
\includegraphics[width=98mm]{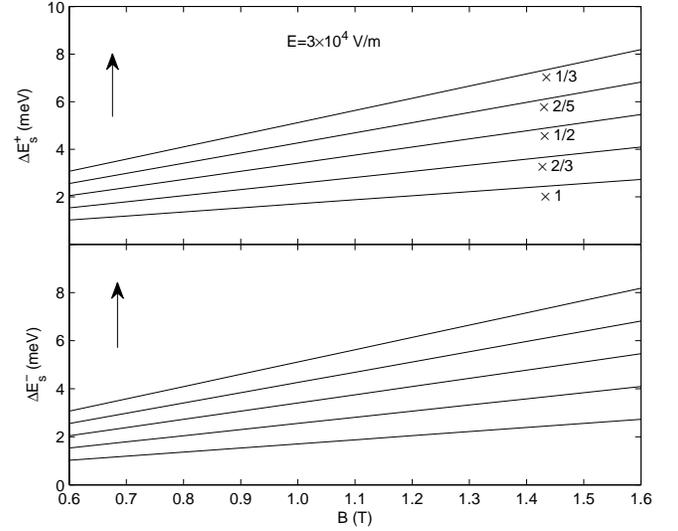}
\caption{Plots of the Landau level spacing in the bulk
versus magnetic field for the fixed electric field. For better 
visualization, each curve is magnified by a suitable constant number as 
indicated in the figure.}
\label{Fig5}
\end{center}
\end{figure}

\section{Numerical result and analysis}
For the numerical calculation, we use the following parameters: 
system size $L_x=L_y=1200$ nm, 
electron's effective mass $m^*=0.068m_e$, 
Rashba spin-orbit coupling constant $\alpha=0.1\alpha_0$ with 
$\alpha_0=10^{-11}$ eV-m.
Selective numerical parameters have been chosen to satisfy the 
following conditions: 
cyclotron radius $R_c \sim l_0 < L_x/2 $ and the shift of the
cyclotron orbit's center due to the maximum applied electric field 
$(x_E)_{\rm max} + x_0 < L_x$.
Here, $(x_E)_{\rm max}=154 $ nm for
maximum $E=5 \times 10^4$ V/m and $B=0.5$ T.
To describe the left and right edge states, we have taken
$k_y = \pm 0.44 L_x/l_0^2$.
Therefore, the left and right edges are far away from each other in
real space. We have also taken $ k_y = 0 $ to describe the bulk states.

\subsection{Spin splitting and Landau level spacing}

In the previous section, we have seen that the spin splitting 
energy and the LLS in the bulk can be controlled by tuning 
the magnetic field. Here, we will see these quantities can also 
be controlled by the transverse electric field.
The spin splitting energy and LLS are plotted in units of 
$\hbar \omega_c$.
In Fig. 1, we show how $\Delta_s$ in the bulk varies with $E$.
Figure 1 shows that the spin splitting energy is diminishing with 
the increase of the applied transverse electric field.
It implies that electric field effectively reduces the effect of 
the RSOI in presence of the perpendicular magnetic field. 
In Fig. 2, we plot $\Delta_s$ at the two edges versus electric
field. 
Comparing Fig. 1 and Fig. 2, it is seen that the spin splitting
energy at the edges is large compared to that of the bulk region
when $E=0$.
The spin splitting energy at the left edge is increasing
with increase of electric field. On the other hand, 
$\Delta_s$ at the right edge is decreasing with increase of $E$.
This is due to the fact that cyclotron orbit shifts from right 
to left with increasing electric field. The states around 
the left edge shifts towards the left boundary where as the 
states around right edge moves towards the bulk region. 
Thats why the spin splitting energy of the left  and right edge 
states increases and decreases with electric field, respectively.
Moreover, the spin-splitting energy at the edges changes 
almost linearly with $E$ whereas $\Delta_s$ in the bulk 
decreases non-linearly with $E$.

In Fig. 3, we show how LLS for spin-up and spin-down branches in
the bulk varies with $E$.
The LLS for upper branches decreases where as for lower 
branches it increases with the electric field.
This is in complete contrast to the case of a 2DES without
RSOI where LLS does not depend on the transverse electric field.
But, it is similar to the LLS in graphene case.
Moreover, for strong enough electric field, 
the LLS for spin-up and spin-down branches tends to
saturate to $\hbar \omega_c$ which is the same as in the
absence of RSOI.
The reduction in the spin splitting energy and 
saturation in LLS in the bulk can be understood from the 
fact that when electric field is sufficiently high
the effective Rashba coupling becomes very weak. 
It is interesting to note that 
$\Delta E_s=\Delta E_s^{+}+\Delta E_s^{-} $ is always $ 2\hbar\omega_c$ 
even in presence of electric field.
We also plot LLS at the edges versus electric field in Fig. 4.

In Fig. 5, we show the LLS as a function of magnetic field when
the transverse electric field is fixed. It shows that both
$\Delta_s^{+}$ and $ \Delta_s^{-}$  increase with magnetic field.
The slope of $\Delta_s^{+}$ is slightly higher than that of the
$ \Delta_s^{-}$.

\begin{figure}[h]
\begin{center}\leavevmode
\includegraphics[width=98mm]{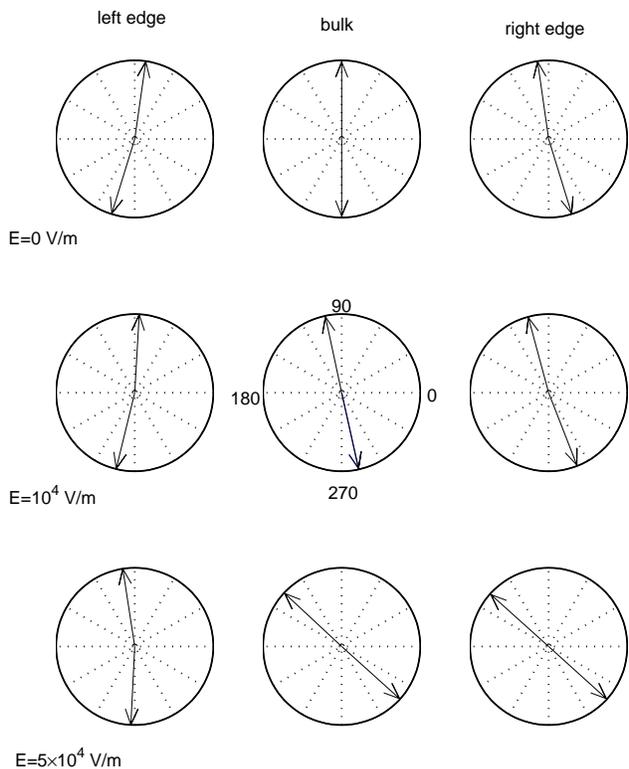}
\caption{Diagrams of the average spin polarization vectors at 
various locations for different electric field strength.}
\label{Fig6}
\end{center}
\end{figure}

\begin{figure}[h]
\begin{center}\leavevmode
\includegraphics[width=98mm]{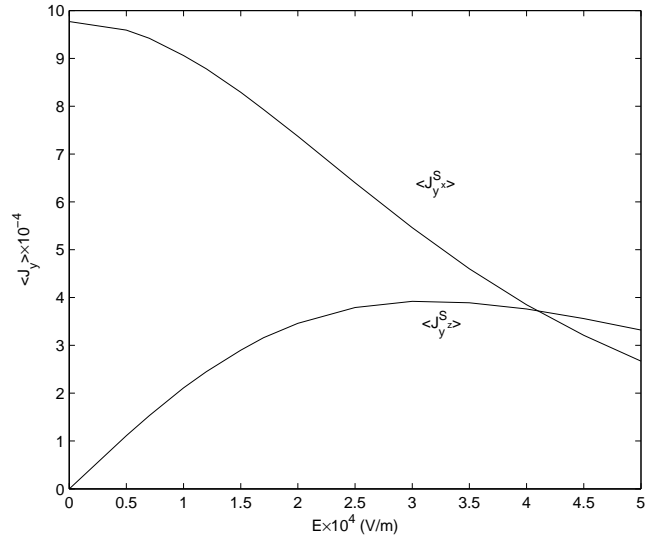}
\caption{Plots of the average spin current for $x$ and $z$ components
versus electric field in the bulk. The average spin current is in units of
$\hbar \omega_c L_x $.}
\label{Fig7}
\end{center}
\end{figure}

\subsection{Average spin polarization}
The average spin polarization vector versus electric field at 
different locations of the center of the cyclotron orbit is shown 
in Fig. 6. The increasing order of rows are accompanied with 
the increasing electric field whereas the columns are 
associated with the three different locations of the cyclotron 
center i.e. left edge, bulk and right edge. The arrow line in 
the upper-half and lower-half circle stands for spin-up and 
spin-down branches, respectively.

The top panel shows the average spin polarization vector in 
absence of the electric field. In the bulk, the average spin 
polarization vector is solely along the $\pm z $ direction 
for spin-up and spin-down branches, respectively. At the edges 
the spin polarization vector lies in the $xz$ plane. 
The magnitudes of the components of the spin polarization
vector depend on the location of the electron.
It is interesting to note that the spin polarization vectors at the
left and rights edges for $+$ branch and $-$ branch are not 
anti-parallel to each other. This is due to strong spin splitting 
at the edges as seen in Fig. 2.   

When a suitable electric field is applied, the average spin 
polarization vectors for up and down states 
start to rotate anti-clockwise. The amount of spin rotation is 
solely determined by the electric field and location. The electric field 
effect on the spin polarization vector at the left edge is
smaller compared to the bulk and right edge. 
This is because of the $k_y$ dependent asymmetric energy spectrum
in presence of electric field.
The anti-clockwise rotation of the average spin-polarization vector
due to the electric field can be observed experimentally 
by Kerr rotation method.

\subsection{Spin current}

Figure 7 shows that in the bulk the average $x$ component 
spin current carried by the Landau level $s = 2$ is destroyed rapidly 
from the maximum value $ \alpha $ due to the electric field, 
whereas the average $z$ component spin current
starts to increase from zero. After certain electric field strength, 
it starts to decrease slowly with the increase of the electric field. 
We have checked that the upper and lower component of the $z$-component 
spin current individually follows the relation between spin polarization 
and spin current i.e; with increasing $x$-component of polarization the 
$z$-component spin current increases.
But the total $z$ component spin current does not follow this after 
a certain electric field strength as shown in the figure.

In Fig. 8, we show how the average $x$ and $z$ components of the 
spin current carried by $s=2$ Landau level change as we increase magnetic 
field for a fixed electric field. The spin polarization along $z$ axis 
increases with increase of the magnetic field. In other words, the spin 
polarization along $x$ axis decreases with increase of the magnetic field.
The average spin current $ J_{s,y}^{\lambda, S_z} $ will decrease with 
magnetic field since $ J_{s,y}^{\lambda, S_z} = - \frac{g \hbar \mu_B B}{2m^*}
P_{s}^{\lambda,S_x}$. Our numerical result is consistent with the exact
analytical results.


\begin{figure}[h]
\begin{center}\leavevmode
\includegraphics[width=98mm]{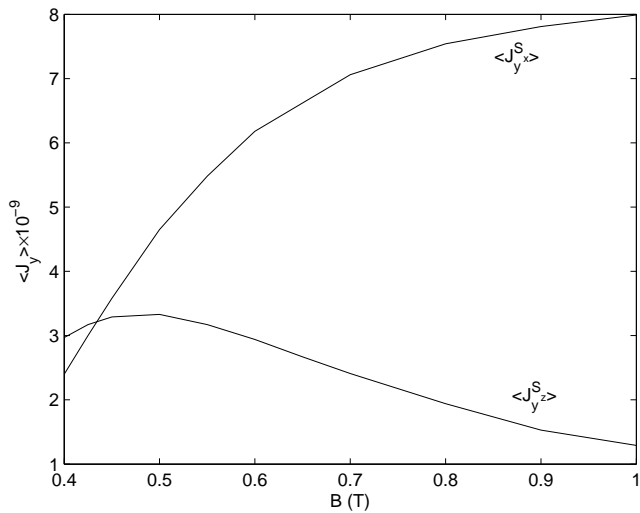}
\caption{Plots of the average spin current for $x$ and $z$ components
versus magnetic field in the bulk. The average spin current is in units of
eV-$L_x $.}
\label{Fig8}
\end{center}
\end{figure}

\section{Conclusion}
We have studied thoroughly the effect of the electric field on the 
2DES with Rashba SOI when the system is under a perpendicular magnetic 
field. The spin splitting energy in the bulk is diminishing with the applied 
transverse electric field.
The Landau level spacing between two successive Landau levels for 
upper or lower branches gets influenced  by electric field. 
This result is in contrast to the 2DES without RSOI, but it is similar 
to the graphene case.
The LLS for upper or lower branches tends to saturate to $\hbar \omega_c $ 
for strong enough electric field. 
These results indicate that strong transverse electric field can 
effectively reduces the effect of the RSOI in presence of magnetic field. 
The electric field also has a strong influence on the spin 
polarization vector depending on the different location of 
electron. The average spin polarization vector rotates 
anti-clockwise as we increase electric field.
The $x$-component of the spin current is destroyed very 
fast by the applied electric field whereas the $z$-component of 
spin current increases initially then decays slowly.
On the other hand, when electric field is fixed and magnetic field
is varying, the $z$-component of the spin current decreases slowly after
a certain magnetic field and $x$-component of the spin current 
increases very fast.

\section{Acknowledgement}
This work was financially supported by the CSIR, Govt. of India under 
the grant CSIR-SRF-09/092(0687) 2009/EMR F-O746.


\begin{thebibliography}{55}

\bibitem{das1}
S. Datta and B. Das,
Appl. Phys. Lett. {\bf 56}, 665 (1990)

\bibitem{appl1}
I. Zutic, J. Fabian, and S. Das Sarma,
Rev. Mod. Phys. {\bf 76}, 323 (2004)

\bibitem{appl2}
A. Wolf et al, 
Science {\bf 294}, 1488 (2001)

\bibitem{appl3}
D. D. Awschalom and M. E. Flatte, 
Nature Phys {\bf 3}, 153 (2007)


\bibitem{tech}
J. Nitta, T. Akazaki, H. Takayanagi, and T. Enoki,
Phys. Rev. Lett. {\bf 78}, 1335 (1997)

\bibitem{matsu}
T. Matsuyama, R. Kursten, C. Meibner, and U. Merkt,
Phys. Rev. B {\bf 61}, 15588 (2000)

\bibitem{rashba}
E. I. Rashba and V. I. Sheka, Fizika Tverdogo Tela; Collected Papers vol
2 (Moscow and Leningrad: Academy of Sciences of the USSR) 162
(1959); E. I. Rashba, Sov. Phys.-Solid State 2, 1109 (1960)


\bibitem{rashba1}
Y. A. Bychkov and E. I. Rashba,
J. Phys. C: Solid State, {\bf 17}, 6039 (1984)


\bibitem{dutta}
B. Das, D. C. Miller, S. Datta, R. Reifenberger,
W. P. Hong, P. K. Bhattachariya, J. Sing, and M. Jaffe,
Phys. Rev. B {\bf 39}, 1411 (1989)

\bibitem{prl99}
J. E. Hirsch, 
Phys. Rev. Lett. {\bf 83}, 1834 (1999)

\bibitem{prl06}
B. A. Bernevig and S. C. Zhang, 
Phys. Rev. Lett. {\bf 96} 106802 (2006)

\bibitem{wang}
X. F. Wang and P. Vasilopoulos, 
Phys. Rev. B {\bf 67}, 085313 (2003)

\bibitem{wang1}
X. F. Wang, P. Vasilopoulos, and F. M. Peeters, 
Phys. Rev. B {\bf 71}, 125301 (2005)

\bibitem{firoz}
SK Firoz Islam and T. K. Ghosh,
J. Phys.: Condens. Matter {\bf 24}, 185303 (2012)

\bibitem{raimondi}
P. Lucignano, R. Raimondi, and A. Tagliacozzo, 
Phys. Rev. B {\bf 78}, 035336 (2008)


\bibitem{kato}
Y. K. Kato, R. C. Myers, A. C. Gossard, and D. D. Awschalom, 
Science {\bf 306}, 1910 (2004)

\bibitem{sinova}
J. Wunderlich, B. Kaestner, J. Sinova, T. Jungwirth, 
Phys. Rev. Lett. {\bf 94}, 047204 (2005)

\bibitem{lukose}
V. Lukose, R. Shankar, and G. Baskaran, 
Phys. Rev. Lett. {\bf 98}, 116802 (2007)


\bibitem{lukose1}
N. M. R. Peres and E. V. Castro,
J. Phys.: Condens. Matter {\bf 19}, 406231 (2007)


\bibitem{ngu}
N. Gu, M. Rudner, A. Young, P. Kim, and L. Levitov,
Phys. Rev. Lett. {\bf 106}, 066601 (2011) 





\bibitem{halperin}
B. I. Halperin, 
Phys. Rev. B {\bf 25}, 2185 (1982)

\bibitem{macdonald}
A. H. Macdonald and P. Streda, 
Phys. Rev. B {\bf 29}, 1616 (1984) 

\bibitem{henrik}
A. Strom, H. Johannesson, and G. I. Japaridze, 
Phys. Rev. Lett. {\bf 104}, 256804 (2010)

\bibitem{muller}
G. Muller, D. Weiss, A. V. Khaetskii, K. von Klitzing,
S. Koch, H. Nickel, W. Schlapp, and R. Losch,
Phys. Rev. B {\bf 45}, 3932 (1992)

\bibitem{chang}
A. M. Chang,
Rev. of Mod. Phys. {\bf 75}, 1449 (2003)

\bibitem{badalyan}
V. L. Grigoryan, A. M. Abiague, and S. M. Badalyan, 
Phys. Rev. B {\bf 80}, 165320 (2009)

\bibitem{rose}
A. Reynoso, G. Usag, M. J. Sanchez, and C. A. Balseiro, 
Phys. Rev. B {\bf 70}, 235344 (2004)



\bibitem{zhang}
Z. Wang, W. Zhang, and P. Zhang, 
Phys. Rev. B {\bf 79}, 235327 (2009)

\bibitem{bao}
Y. Bao, H. Zhuang, S. Shen, and F. Zhang, 
Phys. Rev. B {\bf 72}, 245323 (2005)


\bibitem{sonin}
E. B. Sonin, 
Phys. Rev. B {\bf 76}, 033306 (2007) 




\end{thebibliography}
\end{document}